\documentclass{llncs}

\usepackage{times}
\usepackage{hyperref}       % For the handy \autoref{} command that prevends you from typing Section~\ref{}, also allows for global switching to/from Fig. and Figure
\usepackage{sty/mycommands}

\usepackage[pdftex]{graphicx}

\usepackage{float}

\usepackage{enumitem}
\setlist{nolistsep}

\begin{document}

%\title{Trace Clustering for Very Large Event Data in Healthcare using Frequent Sequence Patterns }
%\title{Trace Clustering in Healthcare using Frequent Sequence Patterns }
\title{Trace Clustering on Very Large Event Data in Healthcare using Frequent Sequence Patterns}

\author{
Xixi Lu\inst{1}
%\orcidID{1111-2222-3333-4444}
\and
Seyed Amin Tabatabaei\inst{2}
%\orcidID{0000-0001-9659-2028} 
\and
Mark Hoogendoorn\inst{2}
%\orcidID{0000-0003-3356-3574}
\and
Hajo A. Reijers\inst{1}
%\orcidID{2222--3333-4444-5555}
}

\institute{
Department of Information and Computing Sciences, Utrecht University\\
\email{\{x.lu, h.a.reijers\}@uu.nl}
\and
Department of Computer Science, Vrije Universiteit Amsterdam \\
\email{\{s.tabatabaei, m.hoogendoorn\}@vu.nl}\\
}

\maketitle
\setcounter{footnote}{0}

\begin{abstract}

Trace clustering has increasingly been applied to find homogenous process executions. % in order to obtain accurate diagnostic information regarding process effectiveness and efficiency. 
However, current techniques have difficulties in finding a meaningful and insightful clustering of patients on the basis of healthcare data. The resulting clusters are often not in line with those of medical experts, nor do the clusters guarantee to help return meaningful process maps of patients' clinical pathways. After all, a single hospital may conduct thousands of distinct activities and generate millions of events per year. 
%of thousands of activities are being recorded per year and unique paths are followed tailored towards patients needs, 
%that is inline with those of medical experts or helps to discover simpler models. 
%
In this paper, we propose a novel trace clustering approach by using sample sets of patients provided by medical experts. 
More specifically, we learn frequent sequence patterns on a sample set, rank each patient based on the patterns, and use an automated approach to determine the corresponding cluster. We find each cluster separately, while the frequent sequence patterns are used to discover a process map.
The approach is implemented in ProM and evaluated using a large data set obtained from a university medical center. 
The evaluation shows F1-scores of 0.7 for grouping kidney injury, 0.9 for diabetes, and 0.64 for head/neck tumor, while the process maps show meaningful behavioral patterns of the clinical pathways of these groups, according to the domain experts. 

\end{abstract}

\section{Introduction}
%\begin{itemize}
%\item
Clinical pathways are known to be enormously complex and flexible. 
%The huge amount of health care event data 
Process mining techniques are often applied to analyze event data related to clinical pathways, in order to obtain valuable insights~\cite{DBLP:journals/jbi/RojasMSC16}. The resulting findings can help to improve process quality, patient outcomes and satisfaction, and optimizing resource planning, usages, and reallocation~\cite{DBLP:journals/cbm/CaronVVLWB14}.
%Such a care process is typically designed for a well-defined patient group and organized to ``standardize the ordering of the activities in a multi-disciplinary team for a specific clinical problem or medical procedure''~\cite{}. In reality, patients often deviate from such a processes. 
Finding coherent, relatively homogenous patient groups helps process mining techniques to obtain accurate insights~\cite{DBLP:journals/tkde/GrecoGPS06,DBLP:conf/bpm/SongGA08,DBLP:journals/tkde/WeerdtBVB13}. 
%
%Finding coherent and homogenous patient groups has been difficult due to the complexity and flexibility. 
%into  clusters can help practitioners to analyze these healthcare processes and to obtain novel insights into such healthcare processes. 
%
%\item

Many existing systems have tried to classify patients and provide such a well-defined group, known as Patient classification systems (PCSs). 
%One domain in which machine learning can play a role is in Patient classification systems (PCSs). 
PCSs provide a categorization of patients based on clinical data (i.e. diagnoses, procedures), demographic data (i.e. age, gender), and resource consumption data (i.e. costs, length of stay)~\cite{schreyogg2006methods}. 
 %The diagnostic related group now serve as a prospective payment system in many countries.
%\draft{Many process mining case studies rely on these codes to find homogenous patient groups.}
%
While useful, such systems often do not align well with the patient groups as clinicians would define them.
%
%\item
Patients who have received the same diagnosis (codes) may be treated for different purposes.
% follow different activities. Similar, the same type of activities may be executed for treating completely different diagnosis. 
%
For example, patients who get reconstructive breast surgery caused by breast cancer or by gender change can be assigned to the same group, while they have different characteristics and should be assigned to different groups, according to medical experts~\cite{aime17amin}. Consequently, the process models derived for such a patient group is often also inaccurate and not aligned with the pathways which clinicians would have in their mind. As a result, much manual work involving medical experts is needed to obtain meaningful patient groups.
%
%Ideally, machine learning techniques would be able to find a suitable grouping that is clinically relevant and useful in practice. Furthermore, the definition of the group itself should be understandable for domain experts.

%\item 
Emerged from the process mining discipline, trace clustering techniques aim to help finding such homogenous groups of process instances (in our case the patients)~\cite{DBLP:journals/tkde/GrecoGPS06,DBLP:conf/bpm/SongGA08,DBLP:conf/bpm/BoseA09a,DBLP:journals/tkde/WeerdtBVB13}.
%\draft{Most exiting approaches either perform pair-wise comparison between the feature vectors of the cases or use a model representation to calculate similarity (dissimilarity) measures between sequences~\cite{DBLP:journals/tkde/WeerdtBVB13}.} 
These techniques cluster the process instances based on the similarity between the sequences of executed activities. 
% trying to discovering more coherent processes. 
%
%\item
%Facing hospitals where 100 thousands of patients per year, thousands of distinct activities conducted, and millions of event recorded, 
However, when applied on hospital data, these approaches face several challenges. Firstly, they have difficulties in scaling-up to handle such large data sets, which may contain hundreds of thousands of patients and millions of events per year. 
Secondly, they assume that the cases within a group show more homogenous behavior than the cases of different groups, whereas in healthcare, 
%99\% of the patients follows a unique path. 
patients treated for the same purpose could have disjoint paths and vice versa. Thirdly, feature vectors (or other intermediate models) are often used to represent the cases and to compute similarity measures; the resulting clusters are often based on an average of the measures and, therefore, may not have clear criteria and may be difficult to explain.
Finally, a resulting cluster of patients could still have thousands of distinct activities, which prevents any process discovery algorithm to find a reasonable process model. 
%and finding a clustering of patients that is meaningful for the domain experts, as shown in~\autoref{fig:Untitled}. 
%As a result, the clusters returned are neither meaningful for the domain experts, nor do they help to discover more structured models that can be shown to or discuss with the experts.  
%\begin{figure}[tb]
	%\centering
		%\includegraphics[width=1.00\textwidth]{figs/problemTC.pdf}
	%\caption{A sketch of the trace clustering problem in healthcare.}
	%\label{fig:Untitled}
%\end{figure}
%\item
%Furthermore, the definition of the group itself should be understandable for domain experts.
%In such settings, validating the clustering with domain experts (such as doctors) is difficult. 

\begin{figure}[tb]
	\centering
		\includegraphics[width=0.9\textwidth]{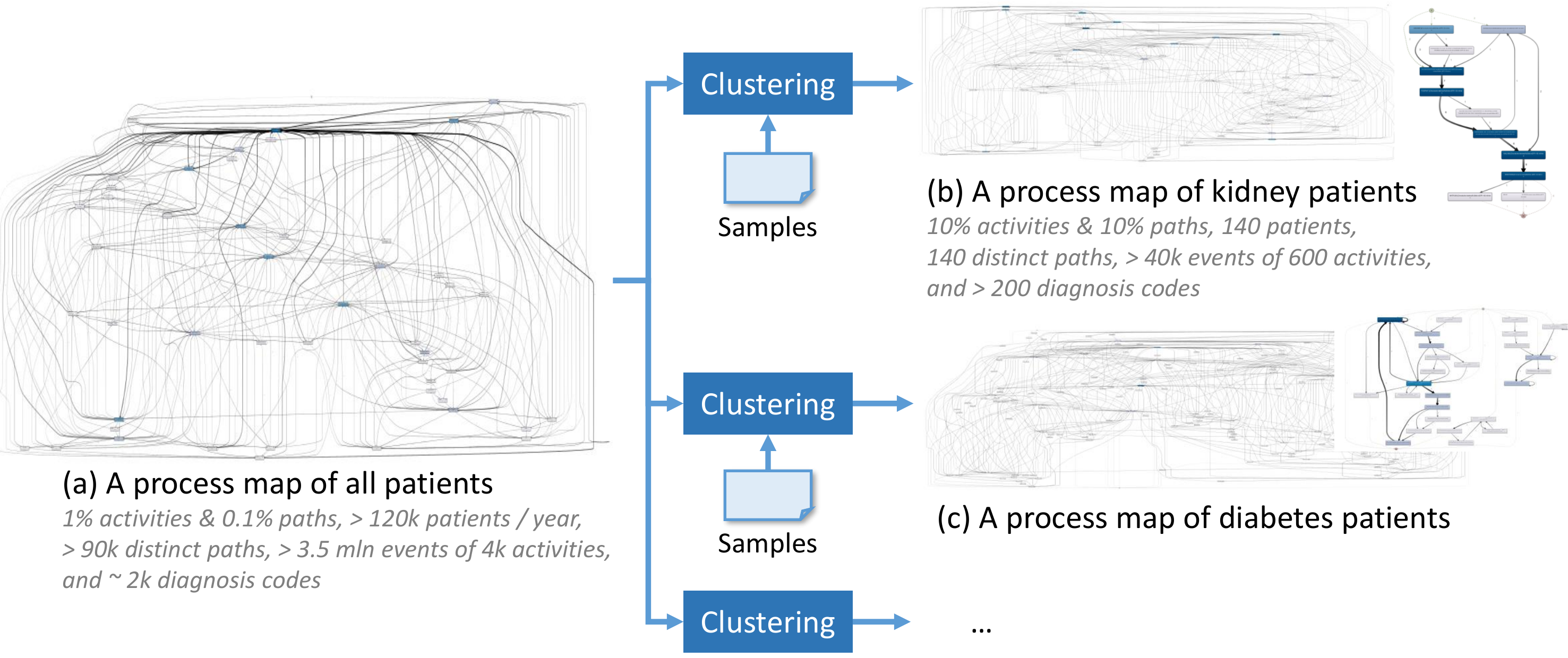}
	\caption{An example of the partial trace clustering problem and an overview of our approach using sample sets and frequent sequence patterns.}
	\label{fig:Untitled2}
\end{figure}

%\item
In this paper, we propose a novel perspective to the trace clustering problem. We use sample sets to find one patient cluster at a time by exploiting frequent sequential pattern mining techniques, exemplified in~\autoref{fig:Untitled2}.
%
%\item
More specifically, \emph{we assume for each group a small sample set of patients (i.e., patient ids) that belongs to the group is made available by medical experts}. 
Using the available event data of patient pathways of the sample set, we compute frequent sequence patterns (FSPs) to learn the behavioral criteria of the group. All patients are ranked based on the behavioral criteria, and we use thresholds to automatically determine whether they belong to the group (see~\autoref{fig:example}, discussed in~\autoref{sec:approach}). Each group is clustered independently. The obtained sequence patterns are used to discover simple process maps.

%number of sequence patterns they match to.
%
%
%\item 
The approach is implemented in the Process Mining toolkit ProM\footnote{\scriptsize\url{http://www.promtools.org/}, in the \emph{TraceClusteringFSM} package (see \href{https://svn.win.tue.nl/repos/prom/Packages/TraceClusteringFSM/Trunk/}{source code})} and evaluated using three real-life cases obtained from a large academic hospital in the Netherlands. The results are validated with a semi-medical expert and a data analyst of the hospital, both of them work closely with medical experts; the semi-medical expert is a manager who has acquired relevant medical knowledge regarding the patient pathways. 
%\draft{The evaluation shows F1-scores of 0.7 for grouping kidney injury and 0.9 for diabetes. Two discovered process maps are shown to and discussed with a domain expert and a data analyst in the hospital who work closely with medical experts. According to the data analyst, the maps show important medical steps of the patient groups and based on her expertise can be used to \todo{}}
% to communicate to the domain experts. 

%\draft{
The contribution of this work is that it gives a concrete method to identify patient clusters from a wealth of data and high variety of pathways with relatively little input from experts. Moreover, this clustering will lead to simple process maps of frequent behavioral patterns in the clinical pathways that can be used in the communication with medical experts. 
Such a method may be useful to reason about clinical pathways within hospitals for the sake of process improvement or quality control.
%}

%\draft{The contribution of this work is that it gives a concrete method to identify patient clusters from a wealth of data with relatively little input from experts. Moreover, this clustering will lead to simple process maps of frequent behavioral patterns of patients. According to the semi-medical expert, they can be a useful tool in the communication with medical experts about the clinical pathways.
%Such a method may be useful to reason about clinical pathways within hospitals for the sake of process improvement or quality control.
%}
%An evaluation of the approach is performed on two real-life cases at
%a large academic hospital in the Netherlands, containing previously identied
%patient groups.

%\item 
In the remainder, we first discuss related work in~\autoref{sec:relatedwork}. 
We recall the concepts and define the research problem in~\autoref{sec:research}
%, the measures used in the evaluation, and the methods used in the approach in~\autoref{sec:prem}.
%We formalize the research problem in~\autoref{sec:research} and explain 
The proposed approach is described in~\autoref{sec:approach}. The evaluation results are presented~\autoref{sec:evaluation}, and \autoref{sec:conclusion} concludes the paper.
%\end{itemize}

\section{Related Work}
\label{sec:relatedwork}
%In this section, we discuss related work while positioning our approach and listing the differences. 
In this section, 
%we discuss trace clustering techniques and other related work. 
%
we discuss three streams of trace clustering techniques, categorizing them by their similarity measures. %An overview of the related work is listed in~\autoref{tab:addlabel}

%We first discuss three types of similarity measures of traces presented in the existing literature. Consequently, we discuss the different types of clustering algorithms. 
%the field of log preprocessing and process discovery and discuss related work while positioning existing techniques using the framework of \autoref{fig:overview}.

%\subsection{Related Work}

%\subsection{Measures of Trace Similarity and Dissimilarity}
%We group the similarity (dissimilarity) measures found in the process mining literature into the following three categories. 

%\begin{itemize}
%\subtopic{Feature vector based similarity}
%\item
\subtopic{Feature-vector-based similarity}
Early work in trace clustering has followed the ideas in traditional data clustering. 
Each trace is transformed into a vector of features based on, for example, the frequency of activities, the frequency of \emph{directly-followed} relations, the resources involved, etc. Between these feature vectors, various distance metrics in data mining are reused to estimate the similarity between the traces. 
%A high distance measure between two traces indicates that the two traces are highly dissimilar. 
Subsequently, distance-based clustering algorithms are deployed, such as k-means or agglomerative hierarchical clustering algorithms~\cite{DBLP:journals/tkde/GrecoGPS06,DBLP:conf/bpm/SongGA08,DBLP:conf/bpm/BoseA09a}. 

In line with feature-vector based trace clustering techniques, 
the work of Greco et al.~\cite{DBLP:journals/tkde/GrecoGPS06} was one of the first approaches that incorporated trace clustering into process discovery algorithms. Their work uses frequent (sub)sequences of activities to constitute feature vectors that represent traces. Hierarchical clusters are then built using a top-down approach which iteratively refines the most imprecise process model (represented as disjunctive workflow schemas).
Song et al.~\cite{DBLP:conf/bpm/SongGA08} present a technique that generalizes the \emph{feature space} by considering data attributes in other dimensions than solely focusing on the control-flow. Features of traces in one dimension are grouped into a so-called \emph{trace-profile}, e.g., resource, performance, case attribute profiles, etc. Furthermore, a multitude of vector-based distance metrics and clustering techniques (both partitioning and hierarchical) are deployed.
In~\cite{DBLP:conf/bpm/BoseA09a}, Bose and van der Aalst compute reoccurring sequences of activities, known as \emph{tandem arrays}, and used these patterns as features in the feature space model in order to improve the way the control-flow information is taken into account in trace clustering.

%\item
%\subtopic{Trace sequence based similarity}
\subtopic{Trace-sequence based clustering}
The second category proposes that the similarity can be measured by the syntax similarity between two trace sequences. 
A trace can be edited into another trace by adding and removing events. The similarity between two traces is measured by the number of the edit operations needed. 
An example of this category of measure is the Levenshtein Edit Distance (LED). 
%\subsection{} 
Bose and van der Aalst~\cite{DBLP:conf/sdm/BoseA09} propose a trace-sequence distance by generalizing the LED and use the agglomerative clustering technique. 
 %The quantified cost is used to express the dissimilarity between traces and build hierarchical clusters by applying an agglomerative clustering technique. 
Chatain et al.~\cite{DBLP:conf/er/ChatainCD17} assume that a normative process model is available and align the traces with the runs of the model. In essence, the traces that are close to the same run (in terms of sequence distance) are clustered into the same group. 

%In~\cite{DBLP:conf/otm/WangHWWHS10}, Wang et al. presents an approach that measures the behavioral similarity between two sequences by computing the longest common subsequence and used this to measure the similarity between the behavioral of two process models.  

%\todo{Alignment based trace clustering}
%\todo{add Chatain's }}

%may also be measures using the number of the operations on the traces as-is. This and compute sequence based similarity metrics between the traces directly, for example LED. 
%One may also reuse existing distance-based clustering algorithms. 
%\item
%\subtopic{Model-based similarity}
\subtopic{Model-based trace clustering} 
Recently, the aim of trace clustering to discover better models has become more prominent. Consequently, the definition of the similarity between traces has shifted from the traces themselves to the quality of the models discovered from those traces. In essence, it is proposed that a trace is more similar to a cluster of traces, if a more fitting, precise, and simple model can be discovered from the cluster~\cite{DBLP:journals/datamine/CadezHMSW03,DBLP:conf/bpm/FerreiraZMF07,DBLP:journals/tkde/WeerdtBVB13}.  

Early work in sequence clustering used first-order Markov models as the intermediate models to represent the clusters. In 2003, Cadez et al.~\cite{DBLP:journals/datamine/CadezHMSW03} proposed to learn a mixture of first-order Markov models from user behavior by applying the Expectation Maximization problem. The approach is evaluated on a web navigation data set. Later, Ferreira et al.~\cite{DBLP:conf/bpm/FerreiraZMF07} followed the same idea and qualitatively evaluated the clustering algorithm in a process mining setting using two additional data sets. 
 
De Weerdt et al.~\cite{DBLP:journals/tkde/WeerdtBVB13} use Petri nets as intermediate models and optimize a F-measure of the models discovered from the clusters. 
The algorithm, called ActiTraC, first samples distinct traces, based on frequency or distance, as initial clusters. The traces that ``fit'' into the intermediate-model of a cluster are assigned to the cluster. The remaining noisy traces either are distributed over the clusters or returned as a garbage cluster. %The objective is to cluster traces in such a way that the intermediate models that are used to represent the clusters are of high quality or satisfy predefined criteria (e.g., recall and precision).\myindex{similarity!model-based}   %A screenshot of the tool is shown in~\autoref{c05fig:clusterActTrac}. 
More recently, De Koninck et al.~\cite{DBLP:conf/caise/KoninckNBBSW17} proposed to incorporate domain knowledge by assuming that a complete clustering solution is provided by experts. The proposed technique then aims to improve the quality of such a complete expert-driven clustering in terms of the model qualities. 

\subtopic{Discussion}
Regarding the feature-based techniques, the number of possible features can be immense, especially in process mining~\cite{DBLP:conf/bpm/SongGA08}. For example, for $n$ activities, we could have $n^2$ number of \emph{directly-followed} relations and $n^3$ if we consider three activities. With thousands of distinct activities, it can be computationally expensive if we consider the full feature space. 
Moreover, as the clusters are calculated based on the average distances between feature vectors, it is often difficult to explain the reason of a particular clustering. 
In many cases, the feature-based techniques have difficulties in finding clusters that are in line with those of domain experts~\cite{DBLP:conf/caise/KoninckNBBSW17}. 
% This also 
Sequence-based trace clustering faces similar limitations as the feature-based. 
Furthermore, patients who have disjoint sets of activities and diagnosis (codes) may belong to the same group. Both feature-based and sequence-based would have difficulties finding those. 
%
%As a result, mostly only a set of the features is selected and computed. Furthermore, because each trace is converted into a feature vector and 
For model-based trace clustering techniques, it would be difficult to handle the complexity of the intermediate models. The clinical pathway of a well-defined patient group could still be extremely complex with thousands distinct activities being executed and each patient following a unique path tailored towards their conditions (see~\autoref{sec:data}). 
Assuming that a complete expert-driven clustering is available would put too much effort on medical experts and is not feasible for this reason. 
%Our approach therefore need to be able to deal with this complexity and find clear criteria for a particular cluster. 
Our approach, therefore, needs to be scalable and able to deal with this complexity.
The approach should also put more emphasis on the abundant domain knowledge available and find clear behavioral criteria of the clusters such that the behavioral criteria are meaningful for domain or medical experts.

\section{Research Problem}
\label{sec:research}
In this section, we first recall the preliminary concepts such as event logs, traces, and activities. Using these concepts, we define our research problem.
\subsection{Preliminaries}
\label{sec:prem}

% Lu ------------------------------------

%\todo{XL: add reference to process mining}

%In this paper, we present an approach to define patients groups based on a small sample of example patients provided by medical experts.

%\subsection{Preliminaries}
A \term{process} describes a set of \emph{activities} executed in a certain order. For example, each patient in a hospital follows a certain \emph{process} to treat a certain diagnosis of a disease, also known as \term{clinical pathway}.  
%Each healthcare process consists of a set of healthcare activities, such as ….
%, such as when and where is this activity executed, and who executed this activity. %Thus, a trace of a patient records a patient treatment trajectory. 
%
An \term{event log} is a set of \emph{traces}, each describing a sequence of \term{events} through the process. Each \emph{event} records additional information regarding the executed \term{activity}. For example, Table \ref{tab:eventloglistexample} shows a snippet of an event log of a healthcare process. Each row records an executed event, which contains information such as the event id, the patient id, the activity, the timestamps, the diagnosis code (also known as diagnosis-related group (DRG)~\cite{schreyogg2006methods}, or DBC in Dutch), and potentially some additional attributes regarding the event.

\begin{table}[tb]
\centering
\caption{An example of an event log of a healthcare process}
\resizebox{.8\textwidth}{!}{
\begin{tabular}{|C{1cm}|C{1.2cm}|L{4cm}|C{2cm}|C{1.2cm}|c|}
%\toprule
\hline
Event & PID  & Activity & Time stamps & \DRG & Attr. \\
\hline
1     & 1001 & Registration (Reg) & 22-10-2018 & \DRG1 &  ... \\
2     & 1001 & Doctor appointment (Doc) & 23-10-2018  & \DRG1    &  ... \\
3     & 1001 & Lab test (Lab) & 24-10-2018     & \DRG1   & ... \\
4     & 1001 & Surgery (Srg) & 30-10-2017     & \DRG2 & ... \\
5     & 1001 & Doctor appointment (Doc) & 01-11-2017 & \DRG2     & ... \\
21    & 1002 & Registration (Reg) & 23-10-2017  & \DRG3 & ... \\
%22    & 1002 & Doctor appointment (Doc) & 24-10-2017 & \DRG3 & ... \\
22    & 1002 & Lab test (Lab) & 25-10-2017  & \DRG3    & ... \\
23    & 1002 & Surgery (Srg) & 26-10-2017  & \DRG4 & ... \\
31    & 1003 & Registration (Reg) & 25-10-2017 & \DRG1 & ... \\
32    & 1003 & Surgery (Srg) & 26-10-2017  & \DRG1 & ... \\
%33    & 1003 & Doctor appointment (Doc) & 26-10-2017 & \DRG1 & ... \\
...     & ...     & ...     & ...     & ...     & ...     \\
\hline
\end{tabular}%
}
\label{tab:eventloglistexample}%
\end{table}%

\begin{definition}[Universes]\label{def:universe}
We write the following notations for universes: 
%\begin{itemize}
%\item 
$\mathcal{E}$ denotes the universe of unique \emph{events}, i.e., the set of all possible event identifiers. 
%\item
$U$ denotes the set of all possible attribute names. 
%\item
$\mathit{Val}$ denotes the set of all possible attribute values.
%\item
$\mathit{Act} \subset \mathit{Val}$ denotes the set of all possible activity names.
%\item
$\mathit{PI} \subset \mathit{Val}$ denotes the set of all possible process instance identifiers. 
%\end{itemize}
\end{definition}

%\begin{definition}[Event]
%Let $A$ be the set of all possible activities and $D$ the set of all possible diagnosis codes. 
%Each event $e_j \in \sigma_i$ is a tuple $(p_j, a_j, d_j, t_j)$ where $a_j \in A$ refers to the activity of $e_j$, and $d_j \in D$ refers the diagnosis code of $e_j$, and $t_j$ is the timestamps of the occurred event $e_j$. 
%\end{definition}

\begin{definition}[Event, Attribute, Label]\label{def:event}
For each event $e \in \mathcal{E}$, for each attribute name $d \in U$, 
%we define the attribute function $\pi_d \in \mathcal{E} \pfun \mi{Val}$ which maps each event $e \in \mathcal{E}$ onto the value assigned to event $e$ for this attribute name $d$. 
the attribute function $\pi_{d}(e)$ returns the value of attribute $d$ of event $e$. 
A labeling function $\pi_{l}: \mathcal{E} \rightarrow \mi{Val}$ is a function that assigns the label to each event $e \in \mathcal{E}$.%\myindex{event}\myindex{event!attribute}\myindex{event!label}
\end{definition}

If the value is undefined, $\pi_{d}(e) = \bot$. Examples of attribute names used in this paper are listed as follows: 
%\begin{itemize}
%\item 
$\pi_{\mathit{pi}}(e) \in \mi{PI}$ denotes the process instance identifier of $e$;
%\item
$\pi_{\mathit{act}}(e) \in \mi{Act}$ is the activity associated with $e$; 
%\item
%$\pi_{l}(e) \in Val$ is the event labeling function that assithe label assigned to $e$; 
%\item
$\pi_{\mathit{time}}(e)$ denotes the timestamp of $e$; 
%\item
$\pi_{\mathit{\drg}}(e)$ denotes the diagnosis code of $e$.
%\end{itemize}
For example, given the log listed in~\autoref{tab:eventloglistexample}, $\pi_{\mathit{pi}}(e_{1}) = 1001$,  $\pi_{\mathit{act}}(e_{1}) =$ Registeration,  $\pi_{\mi{\drg}}(e_{1}) =$ \DRG1.

The labeling function $\pi_{l}(e)$ returns the activity label of event $e$ in the process (also known as an \term{event classifier}). In this paper, we combine both the activities and the diagnosis codes and use them as labels, i.e., $\pi_{l}(e) := \pi_{\mi{act}}(e) + \pi_{\mi{\drg}}(e)$, because the data analyst from the hospital indicated that both are important for the clinical pathway. For example, given the log listed in~\autoref{tab:eventloglistexample}, the label of event 1 is $\pi_{l}(e_1) := \pi_{\mi{act}}(e_1) + \pi_{\mi{\drg}}(e_1) =$ \actlabel{Reg-\DRG1}; the label of event 23 is  $\pi_{l}(e_{23}) =$ \actlabel{Srg-\DRG4}.

%\begin{definition}[Trace, Log]
A \term{trace} $\sigma = \langle e_1, e_2, \cdots, e_n\rangle \in \mathcal{E}^*$ is a sequence of events, where for $1 \leq i < n$, $\pi_{time}(e_i) \leq \pi_{time}(e_{i+1})$ and $\pi_{pi}(e_i)$ =  $\pi_{pi}(e_{i+1})$. An \term{event log} $L = \{ \sigma_1, \cdots, \sigma_{\abs{L}} \} \subseteq \mathcal{E}^*$ is a set of traces.
%\end{definition}
%The executed activities result in such a \emph{trace} of events for the patient, where
%\todo{XL: explain case attributes and example}

%\todo{XL: Which formalization do we use? tuple, function, or other?}

\begin{definition}[Simplified Trace, Simplified Log]
Let $\pi_{l}$ be the labeling function. Let $L = \{\sigma_1, \cdots, \sigma_{\abs{L}}\}$ be a log and $\sigma = \langle e_1, e_2, \cdots, e_{\abs{\sigma}}\rangle$ a trace. We overload the labeling function such that, given $\sigma$, the labeling function returns the sequence of labels of the events in $\sigma$, i.e., $\pi_{l}(\sigma) = \langle \pi_{l}(e_1), \pi_{l}(e_2), \cdots, \pi_{l}(e_{\abs{\sigma}})\rangle \in \mi{Val}^*$. Furthermore, given the log $L$, the labeling function returns the multi-set of the sequences of the labels of the traces in $L$, i.e., $\pi_{l}(L) = [\pi_{l}(\sigma_1), \cdots, \pi_{l}(\sigma_{\abs{L}})]$.

 %$L = \{ \sigma_1, \sigma_2, \cdots, \sigma_m \}$ be the event log, where $\sigma_i = \langle e_1, e_2, \cdots, e_{z_i}\rangle \in L$. 
\end{definition}

Let $\sigma = \langle e_1, \cdots, e_n \rangle \in L$ be a trace. For $1 \leq i < n$, we say event $e_i$ is \emph{directly-followed} by $e_{i+1}$. For $1 \leq i < j \leq n$, we say event $e_i$ is \emph{eventually-followed} by $e_j$. 
For the sake of brevity, we write $L^{l} = \pi_{l}(L)$ and $\sigma^{l} = \pi_{l}(\sigma)$. For instance, the simplified trace of patient 1001 listed in~\autoref{tab:eventloglistexample} is $\pi_{l}(\sigma_{1001}) = \sigma_{1001}^{act,\drg} = \langle $\actlabel{Reg-\DRG1}, \actlabel{Doc-\DRG1}, \actlabel{Lab-\DRG1}, \actlabel{Srg-\DRG2}, \actlabel{Doc-\DRG2}$\rangle$. Note that a patient (an activity) could be associated with multiple diagnosis codes~\cite{schreyogg2006methods}, e.g., patient 1001 (activity \emph{Doc}). 
\subsection{Research Problem - Grouping Patients}

Traditional trace clustering aims to divide the traces of a log into clusters, such that the traces of the same cluster show more homogenous behavior than the traces of different clusters. 
In the healthcare domain, we are facing a very large, complex data set and abundant domain knowledge. As discussed at the end of~\autoref{sec:relatedwork}, 
%, as depicted in~\autoref{fig:Untitled2}. 
%
%\begin{itemize}
%\item
%Firstly, we . %\mi{PI}evertheless, the right number of clusters is still unknown. 
%
%\item
%\draft{It has been difficult to evaluate or to validate the quality of such a large clustering, as the resulting clusters were often not inline with those of medical experts. 
%\item 
%Moreover, the quality of each cluster is highly context-dependent; discussing with a different expert, the patients may be grouped based on different criteria or features~\cite{DBLP:journals/ijmi/BeekGK05}.
%
%For instance, the Dutch patient classification system (DBC) consists of more than two thousands distinct diagnostic codes optimized for grouping patient pathway from a financial perspective~\cite{schreyogg2006methods}. Multiple codes are used for referring to the same disease or the other way around. Experts therefore use additional attributes to define their patient groups of interest~\cite{DBLP:journals/ijmi/BeekGK05}. }
%\end{itemize}
%
we would like to (1) handle such a large data set, to (2) incorporate, leverage, and put more emphasis on the domain knowledge, in order to obtain clusters that are more in line with those of medical experts, while requiring little effort from such experts, and to (3) be able to find the clusters accurately and validate clusters quality, we propose the following. 

We assume that medical experts can provide a small sample set $P$ of the patients that belong to a patient group $\hat{C}$ of interest. %for example a set of patient ids \subset \hat{C}$. 
Giving a sample requires little effort from their side. 
%(1)
%\todo{sum the problems and therefore the solution}
%\textcolor[rgb]{1,0,0}{In such cases where domain knowledge is of important and has specific meanings, we may ask the domain expert to give a sample set. }
%
%Let us assume that there is such a sample set $P$ of the patients given that should be clustered into one group $G$ (i.e., $P \subset G \subset \mi{PI}$). 
%Note that the group $\hat{C}$ is unknown; 
%\draft{
We assume that $\hat{C}$ is unknown (because when $\hat{C}$ gets large, it would require too much effort for medical experts to exhaustively list all patients that belong to $\hat{C}$ and to repeat this process). 
%medical experts, doctors in particular, would be able to exhaustively define the group, but this 
We use the available traces of all patients in the sample $P$, and the objective is to find a cluster $C$ in such a way that $C$ is as close to the group $\hat{C}$ as possible (i.e., the highest recall and precision possible).
We do this separately for each group $\hat{C}_i$ where the sample set $P_i$ is available. 
%
%Let us assume that we have the patient ids $\mi{PI} = \{n_1, \cdots, n_{|\mi{PI}|}\}$, and the event log $L$ of the patients $\mi{PI}$,the set of activities $A$, and the diagnostic codes $D$ as our starting point. Now, a subset $P \subset \mi{PI}$ of the patients that belongs to the group $G$ are provided to us by the domain expert (e.g., the doctor). But the group $G$ is unknown. We would like to compute a set $\leftcenterhat{G} \subset \mi{PI}$ of patients using $P$, such that the set difference between $G$ and $\leftcenterhat{G}$ is minimized. 
%
To generalize, we define the partial trace clustering formally as follows. 

\begin{definition}[Partial Trace Clustering]
%Let $A$ be a set of activities, 
Let $L = \{\sigma_1, \cdots, \sigma_n\}$ be the event log, and $\mi{PI} = \{\pi_{pi}(\sigma_1), \cdots, \pi_{pi}(\sigma_n)\}$ the set of case ids of $L$. Let $P_1, P_2, \cdots, P_x \subset \mi{PI}$ be the sets of samples that respectively belong to clusters $\hat{C}_1, \hat{C}_2, \cdots, \hat{C}_x$, provided by experts (e.g., a doctor), with $x \in \mathbb{N}$. We would like to find the clusters $C_1, C_2, \cdots, C_x \subset \mi{PI}$, such that the set difference between $C_i$ and $\hat{C}_i$ is minimized. 
\end{definition}

Note that clusters $C_1, \cdots, C_x$ can be non-overlapping or form an incomplete clustering of $\mi{PI}$ (i.e., $ C_1 \cup \cdots \cup C_x \subseteq \mi{PI}$), and $x$ could be 1. Based on these properties, we do not have to find all clusters or to compute a complete clustering of all traces. It allows us to mine, cluster, and validate each cluster independently. 

\section{Approach}
\label{sec:approach}
As explained above, we assume that for each cluster $C$ to be found we have a small sample $P$ of the cases that belongs to the true-but-unknown cluster $\hat{C}$. For all other cases it is unknown whether they belong to the cluster or not. By exploiting the available sample set $P$ and the event log $L$ of all cases, the objective is to find \emph{behavioral criteria} for determining the cluster. To find the behavioral criteria and to handle the large number of features, we compute frequent \emph{behavioral patterns}.
%
%To handle the large number of features and to find behavioral patterns, 
In~\autoref{sec:findfsp}, we first explain the use of sequence pattern mining to learn the frequent sequence patterns (FSPs) of the sample set. 
%
%As the frequent sequence patterns reflect frequent behavioral patterns of the process executions, we use them as the behavioral criteria for the group. 
%
In~\autoref{sec:ranktraces}, we then match the FSPs to the other cases in the sample to train our parameters. Finally, we match all cases to the clustering criteria to return the computed cluster in~\autoref{sec:determinecriteria}. \autoref{fig:example} shows an overview of the approach.

\subsection{Finding Frequent Sequence Patterns}\label{sec:findfsp}
The first step of the approach is to find frequent sequence patterns repeated among the samples. A \term{frequent sequence pattern} is a sequence that occurs in the traces with a frequency no less than a specified threshold. We adapt the definition of sequence patterns in our context as follows.  

\begin{definition}[Sequence Pattern]
A sequence pattern $\sp = \langle a_1, \cdots, a_m \rangle \in Val^*$ is a sequence of labels in which $a_i$ is said to be eventually followed by $a_{i+1}$ for $1 \leq i < m$. 
\end{definition}

When a trace \emph{matches} a sequence pattern, it means that the trace contains a sub sequence where the labels occur in the same order. 

\begin{definition}[Support of Sequence Pattern]
Let $L$ be an event log and $\pi_{l}$ the labeling function. 
Let $\sigma \in L$ be a trace, with $\pi_{l}(\sigma) = \langle a_1, a_2, \cdots, a_n \rangle$. 
Let $\vec{s} = \langle s_1, \cdots, s_m \rangle \in \mi{Val}^*$ be a sequence pattern.
We say $\sigma$ \emph{matches} $\sp$ if and only if there exist integers $i_1, i_2, \cdots, i_m$ such that $1 \leq i_1 < i_2 < \cdots < i_m \leq n$ and 
%integers $i1 i2,...,im exist, such that
$s_1 = a_{i_1} , s_2 = a_{i_2} ,..., s_m = a_{i_m}$. We write $\sp \sqsubseteq \pi_{l}(\sigma)$. %super-sequence of $\alpha$ or that 

The \term{support} of sequence $\sp$ in $L$ is the number of traces in $L$ that matches $\sp$, i.e., 
%\[
%\mi{supp}(\sp, L^{l}) = \frac{\abs{\{\sp \sqsubseteq \sigma^{l} \mid \sigma^{l} \in L^{l}\}}}{\abs{L^{l}}}
%\] 
\[
\mi{supp}(\sp, L) = \frac{\abs{\{\sp \sqsubseteq \pi_{l}(\sigma) \mid \sigma \in L\}}}{\abs{L}}
\] 
\end{definition}

Let $\phi_s$ denote the minimum support threshold. A sequence pattern $\sp$ is said to be \emph{frequent} if and only if $\mi{supp}(\sp, L) \geq \phi_s$. We write $\mi{SP}(L, \phi_s)$ to denote the set of all sequence patterns in $L$ with a support of at least $\phi_s$, i.e., 
\[
\mi{SP}(L, \phi_s) = \{ \sp \in Val^* \mid \mi{supp}(\sp, L) \geq \phi_s\}
\]

\begin{figure}[tb]
	\centering
		\includegraphics[width=1.00\textwidth]{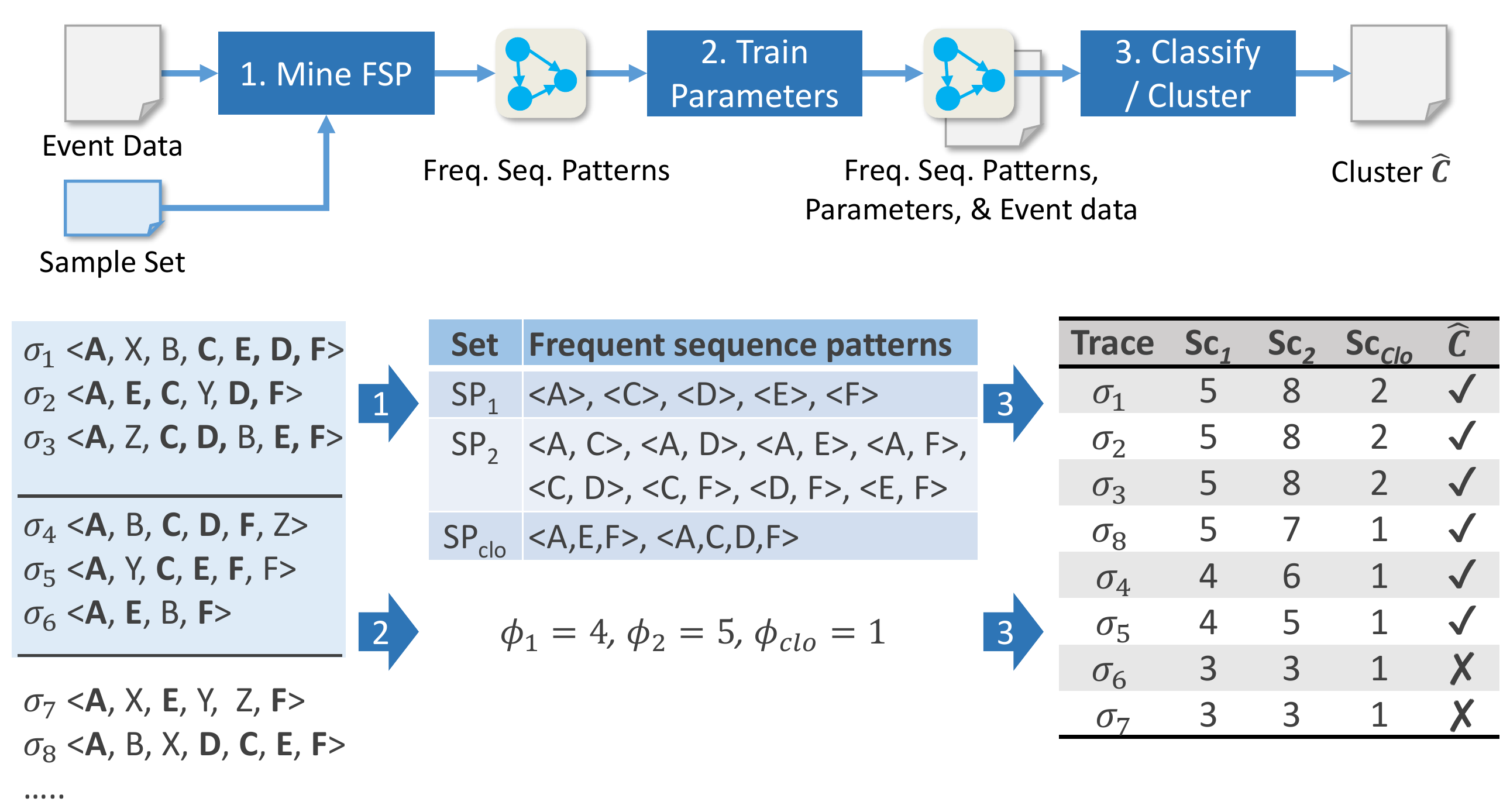}
	\caption{An example of our approach applied on the event log (on the left), with the FSPs mined (in the middle), and the scores of the traces (on the right).}
	\label{fig:example}
\end{figure}

Step 1 in \autoref{fig:example} exemplifies mining frequent sequence patterns, with the minimum support $\phi_s = 0.8$. Let $L' = \{\sigma_1, \sigma_2, \sigma_3\}$, as shown in~\autoref{fig:example}. We have $\mi{SP}(L', 0.8) = \{ \trbr{A}, \trbr{C}, \trbr{D}, \trbr{E}, \trbr{F}, \trbr{A,C}, \trbr{A,D}, \trbr{A,E}, \cdots, \trbr{A, C, D, F} \}$. 
%\todo{explain}

There are several well-known algorithms to compute frequent sequence patterns. 
In this paper, we use the CloFAST algorithm~\cite{DBLP:journals/kais/FumarolaLCM16} and its SPMF implementation~\cite{DBLP:journals/jmlr/Fournier-VigerG14} due to its fast run-time, which is also used in~\cite{DBLP:conf/ideas/CeciSLM18} for next activity prediction. 
%It uses a data structure known as \emph{sparse id-list} (SIL), which largely decreased the run time of computing sequence patterns~\cite{DBLP:journals/kais/FumarolaLCM16}.

\subsection{Trace Ranking By Sequence Pattern Matching}\label{sec:ranktraces}

To automatically find behavioral criteria for determining the cluster $C$, 
we divide the sample set $P$ into a training set $P_{tr}$ and use the entire $P$ as our test set. 
On the training set $P_{tr}$, we compute the set of frequent sequence patterns (FSPs). 
%Finally, the parameters are set based on the estimated recall. 

%First, let $L_{ptr}$ denote the log of traces of patients in $P_{tr}$, i.e., $L_{ptr} = \{\sigma \in L \mid \pi_{pi}(\sigma) \in P_{tr}\}$. We compute the set of frequent sequence patterns as $SP(L_{ptr}, \phi_s)$. 
The FSPs mined on the training set $P_{tr}$ could still be very large. Therefore, we select a subset of the FSPs. We use the FSPs of length 1, 2, and the closed sequence patterns as our behavioral criteria. Note that this step can be generalized with ease and any other subset of the FSPs can be selected as behavioral criteria. We select these three subsets because of the following. The FSPs of length 1 represent the frequent activity labels occurred in the training set; the FSPs of length 2 represent the frequent \emph{eventually-followed} relations occurred. 
The \emph{closed sequence patterns} $\sp \in \mi{SP}(L, \phi_s)$ are the sequence patterns $\sp$ such that for all other patterns which $\sp$ satisfies have a lower support. Thus, these three provide a good coverage of all FSPs with less redundancy.

\begin{definition}[Closed Sequence Pattern]
Let $\mi{SP}(L, \phi_s)$ denote all frequent sequence patterns in $L$ with a support of at least $\phi_s$. A sequence pattern $\sp \in \mi{SP}(L, \phi_s)$ is a closed sequence pattern if and only if for all $\sp'\in \mi{SP}(L, \phi_s)$, $\sp \sqsubseteq \sp' \Leftrightarrow (\sp = \sp' \vee supp(\sp, L) > supp(\sp', L))$. 
\end{definition}

%\[
%\mi{SP}_{clo}(L, \phi) = \{ \sp \in \mi{supp}(\sp, L) \mid  \geq \forall \sp'\in \mi{supp}(\sp, L),   \}
%\]

Next, using these subsets of these patterns, we rank each trace in $P$ based on the number of patterns it satisfies. 
Let $\mi{SP}(P, \phi_s) = \{\sp_1, \sp_2, \cdots \sp_n\}$ be the set of frequent sequence patterns of $P$ above support threshold $\phi_s$. 
Let $\mi{SP}_1, \mi{SP}_2, \mi{SP}_{\mi{clo}}\subseteq \mi{SP}(P, \phi_s)$ be the set of patterns of size 1, size 2, and closed sequence patterns, respectively. We give each case a score based on the number of patterns in $\mi{SP}_1, \mi{SP}_2$, and $\mi{SP}_{\mi{clo}}$ the trace satisfies and rank the cases based on their score. Thus, 

\[
\mi{score}_k(\sigma) = \abs{\{\sp \in \mi{SP}_k | \sp  \sqsubseteq \pi_{l}(\sigma) \}}
\]

For example, see \autoref{fig:example}, step 1 shows $\mi{SP}_1$ of five sequence patterns, $\mi{SP}_2$ of eight, and $\mi{SP}_{\mi{clo}}$ of two, which are mined on the $P_{\mi{tr}} = \{\sigma_1, \sigma_2, \sigma_3\}$ using a minimum support of 0.8. Given trace $\sigma_4 \notin P_{\mi{tr}}$, it matches to $\langle A \rangle$, $\langle C \rangle$, $\langle D \rangle$, and $\langle E \rangle$ in $\mi{SP}_1$, to $\langle A, C \rangle$, $\langle A, D \rangle$, $\langle A, F \rangle$, $\langle C, D\rangle$, $\langle C, F\rangle$, and $\langle D, F\rangle$ in $\mi{SP}_2$, and to $\langle A, C, D, F\rangle$ in $\mi{SP}_\mi{clo}$. Thus, $\mi{score}_1(\sigma_4) = 4$, $\mi{score}_2(\sigma_4) = 6$, $\mi{score}_\mi{clo}(\sigma_4) = 1$.

\subsection{Computing Criteria Threshold}\label{sec:determinecriteria}

%Following our previous work~\cite{}, we
For each case, we have now computed $\mi{score}_1$, $\mi{score}_2$, and $\mi{score}_\mi{clo}$, as explained above. For the three scores, we respectively introduce three thresholds, $\phi_1 \in \mathbb{N}$, $\phi_2\in \mathbb{N}_0$, and $\phi_\mi{clo} \in \mathbb{N}_0$. 
We decide on whether a case belongs to cluster $C$ based on whether the scores of the trace are above the corresponding thresholds, i.e., 

\[
C_{\phi_{1}, \phi_{2}, \phi_\mi{clo}} = \{\pi_{pi}(\sigma)| \sigma \in L \wedge \mi{score}_1(\sigma) \geq \phi_1 \wedge 
\mi{score}_2(\sigma) \geq \phi_2 \wedge 
\mi{score}_\mi{clo}(\sigma) \geq \phi_\mi{clo} \}
\]
%The former is affected by the support threshold ($\phi_s$) we set, the lower the support threshold the more \FSP{s} we find. The latter depends on the number of sequence patterns found. 

%Given a log $L$, a sample $P \in PI$, and the frequent patterns $SP(L, \phi_s)$. 
%Let $\phi_{1}, \phi_{2}, \phi_{clo}$ denote the thresholds for $score_1$, $score_2$, and $score_{clo}$, respectively. We return a trace $\sigma$ as a case in $C$ if and only if all its scores are greater or equal than the threshold:

To estimate the quality of $C_{\phi_{1}, \phi_{2}, \phi_\mi{clo}}$, we then compute the estimated recall with respect to $P$, i.e., 
%\[
$
\overline{\mi{recall}}_{\phi_{1}, \phi_{2}, \phi_\mi{clo}} 
= \frac{\abs{C_{\phi_{1}, \phi_{2}, \phi_\mi{clo}}  \cap P}}{\abs{P}}$. 
%\]
When the sample set $P$ gets closer to the ideal cluster $\hat{C}$, the estimated $\overline{\mi{recall}}$ gets closer to the true recall. When we decrease $\phi_{1}, \phi_{2},$ and $\phi_\mi{clo}$, more cases are included in $C$. After a certain point, the increase of $\overline{\mi{recall}}$ starts to flatten, which suggests that further lowering the thresholds does not help to retrieve a large number of true positive cases, which is likely to result in a low precision. 
To approximate such a point, we use 
%\[
$\mi{max}_{\phi_{1},  \phi_{2}, \phi_\mi{clo}}
%1\leq \phi_{1}\leq \abs{\mi{SP}_1}, 
%0\leq \phi_{2}\leq \abs{\mi{SP}_2}, 
%0 \leq \phi_\mi{clo} \leq \abs{\mi{SP}_\mi{clo}}} 
\frac{\overline{\mi{recall}}^2_{\phi_{1}, \phi_{2}, \phi_\mi{clo}}}{ \abs{{C}_{\phi_{1}, \phi_{2}, \phi_\mi{clo}}}}
%= \frac{\abs{C_{\phi_{1}, \phi_{2}, \phi_{clo}}  \cap P}}{\abs{P}}
$
%\]
%$\frac{\overline{\mi{recall}}^2}{\abs{C}}$ 
~\cite{lee2003learning}, but only consider the thresholds when $\overline{\mi{recall}} \geq 0.8$. The number of iterations to find such a maximum depends on the maximal values of $\mi{score}_1$, $\mi{score}_2$, and $\mi{score}_\mi{clo}$.

\section{Evaluation}
\label{sec:evaluation}
We implemented the described approach in the process mining toolkit ProM. We used a real-life data set to evaluate our approach with respect to the following three objectives: 
%\subsection{Objectives}
\begin{enumerate}[leftmargin=2\parindent]
\item
[(EO1)] How accurate (in terms of F1-scores) are the clusters returned by our automated approach, compared to the optimal scores? 
\item
[(EO2)] How accurate can we find the clusters using our approach, compared to a related approach that uses frequent item sets (FIS)~\cite{aime17amin}?
%\item
\item
[(EO3)] Can we discover a simple and insightful behavioral criteria for each patient group such that the criteria can be used to communicate with medical experts?
\end{enumerate}
In the following, we first discuss the data set in~\autoref{sec:data} and then report our results~\autoref{sec:results} with respect to these three objectives. 
All experiments are run on an Intel Core i7-
8550U 1.80GHZ with a processing unit of 16GB running Windows 10 Enterprise. The maximal queue size of CloFAST algorithm~\cite{DBLP:journals/kais/FumarolaLCM16,DBLP:journals/jmlr/Fournier-VigerG14} is set to $10^5$.   
The obtained results were discussed with the semi-medical expert and the data expert in the hospital who cooperate closely with medical experts in their daily work.

%15597235

%Figure 1a shows the number of patients and Figure 1b shows the number of activities in different years. 

\subsection{Experimental Setup}\label{sec:data}
For the evaluation, we used anonymized patient records provided by the VU University Medical Center Amsterdam, a large academic hospital in the Netherlands. 
%\todo{make the experts explicit..}
All patients that have a diagnosis code registered between 2013 and 2017 are selected. The administrative and dummy activities are filtered out. As a result, we have in total 328,256 patients over the five years. There are 7,426 unique activities and 2,251 unique diagnosis codes registered. In total more than 15.5 million events are recorded in the logs.

% Table generated by Excel2LaTeX from sheet 'Sheet1'
\begin{table}[tb]
  \centering
  \caption{General information of the real-life data set and the ground truth clusters.}
	  \resizebox{.9\textwidth}{!}{
    \begin{tabular}{l|r|r|R{0.8cm}|R{1.4cm}|R{0.8cm}|R{0.8cm}|r|r|R{1.2cm}|R{1cm}}
		%\begin{tabular}{lrrR{0.8cm}rR{0.8cm}R{0.8cm}rrR{1.2cm}R{1cm}}
    \toprule
       Data   & \#cases & \#dpi & \#avg. c/dpi & \#events & \#avg. e/c & \#max. e/c & \#acts & \#dbcs & \#dst. labels & Perc. of all \\
    \midrule
    All17 & 128,505 & 97,771 & 1.3   & 3.70$*10^6$ & 28.8  & 2,924 & 4,666 & 1,915 & 150,244 & - \\ %3,698,124
    \midrule
    $\hat{C}$\_Kidney17 & 140   & 140   & 1.0   & 40,071 & 286.2 & 2,167 & 676   & 237   & 4,777 & 0.11\% \\
    $\hat{C}$\_Diabetes17 & 1,521 & 1,520 & 1.0   & 139,454 & 91.7  & 2,861 & 1,414 & 646   & 16,496 & 1.18\% \\
    $\hat{C}$\_HNTumor17 & 1,050 & 1,048 & 1.0   & 105,613 & 100.6 & 905   & 1,001 & 380   & 9,211 & 0.82\% \\
		    \midrule
		    \multicolumn{11}{c}{...} \\
    \midrule
    All13 & 133,438 & 99,196 & 1.3   & 4.32$*10^6$ & 32.4  & 2,558 & 4,871 & 1,813 & 168,096 & - \\ %4,318,077
		    \midrule
    $\hat{C}$\_Kidney13 & 81    & 81    & 1.0   & 26,949 & 332.7 & 1,577 & 651   & 159   & 3,601 & 0.06\% \\
    $\hat{C}$\_Diabetes13 & 1,573 & 1,573 & 1.0   & 142,737 & 90.7  & 1,057 & 1,427 & 663   & 16,966 & 1.18\% \\
    $\hat{C}$\_HNTumor13 & 1,350 & 1,334 & 1.0   & 147,491 & 109.3 & 2,227 & 1,237 & 437   & 12,678 & 1.01\% \\
    \bottomrule
    \end{tabular}%
		}
  \label{tab:allpatients}%
	
	%\scriptsize{the number of cases, the number of distinct process instances (dpi); the average number of cases per dpi; the number of events, the average number of events per case, the maximal number of events per trace, the number of distinct activities, the number of distinct dbcs, the number of distinct labels, and the number of traces in the cluster as the percentage of the total number of traces}
\end{table}%

In addition, lists of patients of three groups divided over the five years are provided by the analyst, patients with kidney failure, with diabetes, or with head/neck-tumor. We use $\hat{C}_{KidneyYY}$, $\hat{C}_{DiabetesYY}$, and $\hat{C}_{HNTumorYY}$ to refer to them, respectively, where $YY$ denotes the particular year. \autoref{tab:allpatients} lists the number of cases (c), distinct process instances (dpi), events (e), activities (acts), and other statistical information related to the event logs of 2013 and 2017 as examples. 
For instance, in \autoref{tab:allpatients} row 2, 3, and 4 show an overview of $\hat{C}_{Kidney17}$, $\hat{C}_{Diabetes17}$, and $\hat{C}_{HNTumor17}$, respectively. These 15 clusters are used as the ground truth. For each cluster, 30 patient ids are provided by medical experts as the sample set (i.e., $\mid \!\! P \!\!\mid\, = 30$), the same as a previous study~\cite{aime17amin}.  
%We then 
For finding the clusters, we use all the patient records of the same year and the provided $P$ to compute our cluster $C$. The quality of $C$ is evaluated against the corresponding ground truth cluster $\hat{C}$ by calculating the recall, precision, and F1-score, i.e., 
%\begin{align}
%\[
$recall(C, \hat{C}) = \frac{\abs{C \cap \hat{C}}}{\abs{\hat{C}}}$, 
%\end{align}
%\]
%\begin{align}
%\[
$precision(C, \hat{C}) = \frac{\abs{C \cap \hat{C}}}{\abs{C}}$, and 
%\end{align}
%\]
%
%
%\begin{align}
%\[
$F1\_measure(C, \hat{C}) = 2 \cdot  \frac{precision(C, \hat{C}) \cdot recall(C, \hat{C})}{precision(C, \hat{C}) + recall(C, \hat{C})}$. 
%\end{align}
%\]
%\draft{

%The obtained results were discussed with a domain expert and a data expert in the hospital who work closely with medical experts. 
%According to the data analyst, the maps show important medical steps of the patient groups and based on her expertise can be used to \todo{}}

It is worthwhile to mention that the data analyst stated that it took a lot of time and effort to obtain each of these ground truth clusters. Multiple intensive discussion sessions were scheduled with different groups of medical experts to come to the definitions and criteria for each of these clusters. This suggests that if our algorithmic approach can identify the behavioral criteria and the clusters with a reasonable accuracy using only a small sample set, it would help reducing the workload of both the analysts and the medial experts and making this process feasible to be repeated for other patient clusters or in other hospitals. 
Another important remark is that the ground-truth clusters which we are trying to find are very small and unbalanced compared to the full event logs, making this trace clustering problem a very challenging task. For instance, the $\hat{C}$\_Kidney17 ($\hat{C}$\_Diabetes17) contains only 140 (1521) patients, about 0.1\% (1.2\%) of the 128,505 patients in the log of 2017.

%%\begin{align}
%\[
%recall(C_{groupYY}, \hat{C}_{groupYY}) = \frac{\abs{C_{groupYY} \cap \hat{C}_{groupYY}}}
%{\abs{\hat{C}_{groupYY}}}
%%\end{align}
%\]
%
%%\begin{align}
%\[
%precision(C_{groupYY}, \hat{C}_{groupYY}) = \frac{\abs{C_{groupYY} \cap \hat{C}_{groupYY}}}{\abs{C_{groupYY}}}
%%\end{align}
%\]
%%
%
%%
%%\begin{align}
%\[
%F1\_measure(C_{groupYY}, \hat{C}_{groupYY}) = 2 \cdot  \frac{precision(C_{groupYY}, \hat{C}_{groupYY}) \cdot recall(C_{groupYY}, \hat{C}_{groupYY})}
%{(precision(C_{groupYY}, \hat{C}_{groupYY}) + recall(C_{groupYY}, \hat{C}_{groupYY})}
%%\end{align}
%\]

\subsection{Results}\label{sec:results}

%\subsubsection{RQ0. Frequent Sequence Patterns}

\subsubsection{EO1) F1 scores of automated approach compared to maximum.}

To determine the support threshold $\phi_s$, we started with 1.0 and decreased the value by 0.1 until a reasonable large amount of patterns are found and the F1 scores stopped increasing. 
For the $C\_\mi{Kidney}$ groups, $\phi_s$ ranges from 1.0 down-to 0.6, 
for $C\_\mi{Diabetes}$, 0.4 down-to 0.2, and for $C\_\mi{HNTumor}$, 0.5 down-to 0.2.
We used either 10 or 15 (of the 30 in the sample) as the training set to learn the frequent sequential patterns, i.e., $k = \, \mid \!\! {P_{\mi{tr}}}  \!\!\mid\, \in \{10, 15\}$.  
We write \TCa for our approach with the automatically determined $\phi_1$, $\phi_2$, and $\phi_\mi{clo}$; \TCb for the maximum F1 score using the same $\phi_s$ and $k$ but based on the optimal $\phi_1, \phi_2,$ and $\phi_\mi{clo}$. 

\begin{figure}[tb]
	\centering
		%\begin{minipage}{.5\textwidth}
		\includegraphics[width=0.98\textwidth]{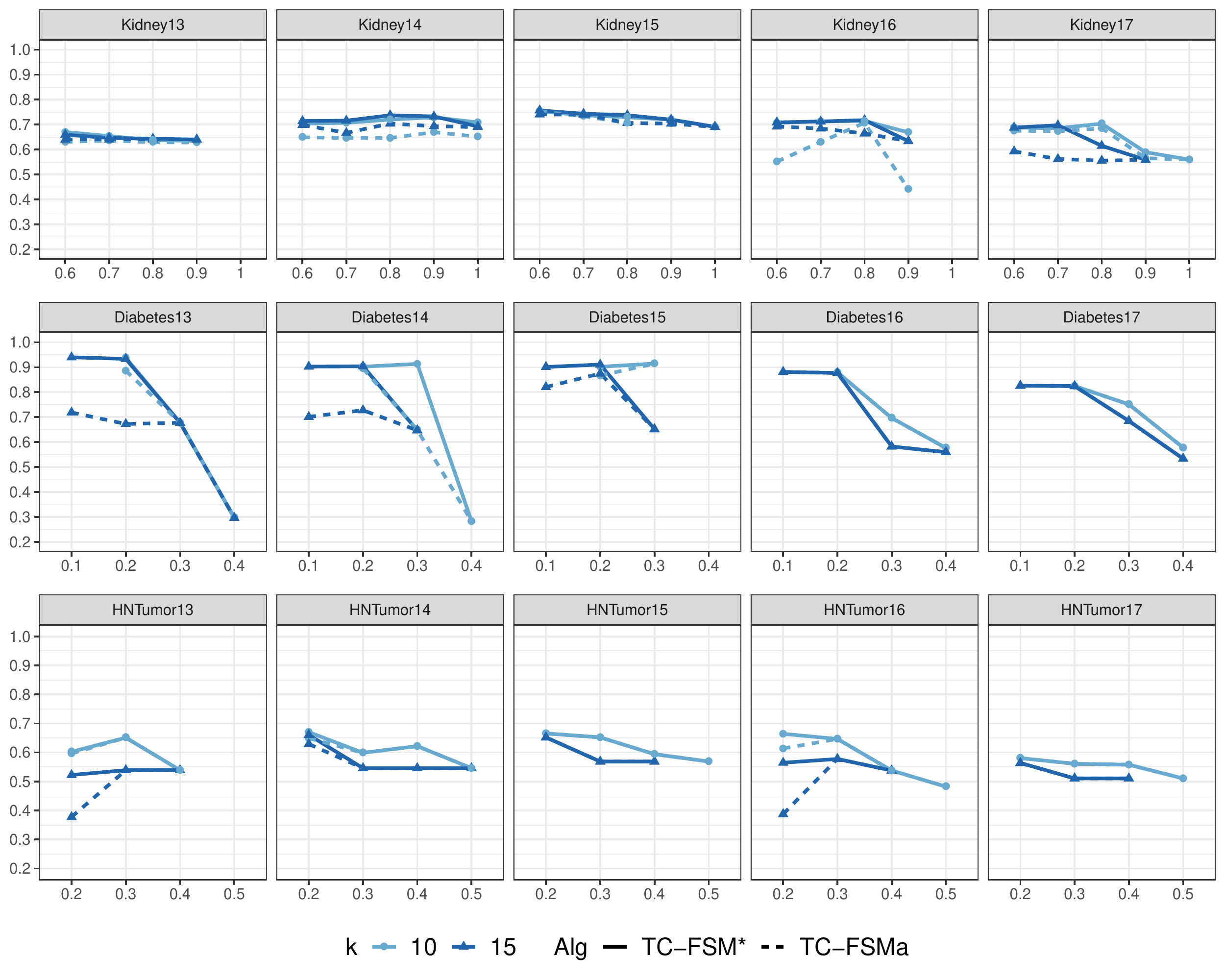}
	\caption{The differences in the F1-scores of the automated approach (\TCa) shown in dotted lines and the maximum scores achieved (\TCb) shown in filled lines, using various support threshold (on the x-axis) and training sample size $k$. }
	\label{fig:res2}
	%\end{minipage}%
\end{figure}

\autoref{fig:res2} shows the difference in F1-scores between \TCa (dotted lines) and \TCb (filled lines). 
We observe that in most cases the F1-scores of the automated \TCa (dotted line) are very close to the ones of the optimal \TCb (filled line). For some clusters, for example Diabetes16\&17 and HNTumor15\&17, \TCa returns the exact same F1-scores as the maximum for all $\phi_s$ and $k$. Only in a few cases, for example, for Diabetes13 when $k$ is 15 and the support $\phi_s$ is 0.2, \TCa scores considerably lower than \TCb with a difference of 0.26. Nevertheless, for the same $\phi_s$ when $k$ is set to 10, this difference is immediately decreased to 0.01.
Taking into account that we only have a sample set of 30 patients and the number of activities ranges in the thousands, the \TCa is able to approximate the optimal F1-scores very well.

For the support threshold $\phi_s$, we observe overall a slight increase in the F1-scores for the Kidney and HNTumor groups when we decrease $\phi_s$. For the Diabetes groups, there is a considerable increase in F1-scores during the beginning (when $\phi_s$ is decreased from 0.4 to 0.2), but this improvement also fades out. One reason for this is because when the support threshold is low, more patterns are found and used as criteria; thus, more patients are included in the cluster including false positives. While the recall increases, the precision becomes lower, which led to a small increase in the F1-scores. For the diabetes group, when the support $\phi_s$ is 0.4, the number of sequence patterns is extremely small (1 or 3). When the support decreases, it allowed the algorithm to find a consider number of defining patterns that is significant to retrieve the patients of the ground truth clusters. This increases the recall dramatically without a significant decrease in precision. 
Furthermore, \autoref{fig:res2} also shows that using fewer training samples ($k = 10$, denoted using light blue), our approach can achieve the same scores as when using a larger training sample set ($k = 15$, denoted using darker blue). In many cases, the former (i.e., $k=10$) even achieved a better result. 
This may be due to that the training test set $P\backslash P_{tr}$ is larger.

%\todo{refine and explain}

%TRAINING SIZE COULD BE LOW

%\begin{figure}[tb]
	%%\begin{minipage}{.5\textwidth}
	%\centering
		%\includegraphics[width=1.00\textwidth]{figs/res3.pdf}
	%\caption{The differences in the F1-scores returned by the automated approach \TCa and the maximum recall achieved. (TODO: update the numbers).}
	%\label{fig:res3}
		%%\end{minipage}
%\end{figure}

\subsubsection{EO2) Comparing the F1-scores achieved.}

We write \TCb to refer to the maximum score of our approach using the above settings. We write \TCs to denote our approach with a single setting (0.8 for kidney, 0.2 for diabetes, and 0.2 for NH-tumor, with sample size 30 and $k_{tr} = 10$), to compare our results to the previous work~\cite{aime17amin}. The parameters are selected on the results of EO1. We write \FIS for referring to the previous approach that uses frequent item sets~\cite{aime17amin}.

\begin{figure}[tb]
	\centering
		\includegraphics[width=.9\textwidth, height=4.6cm]{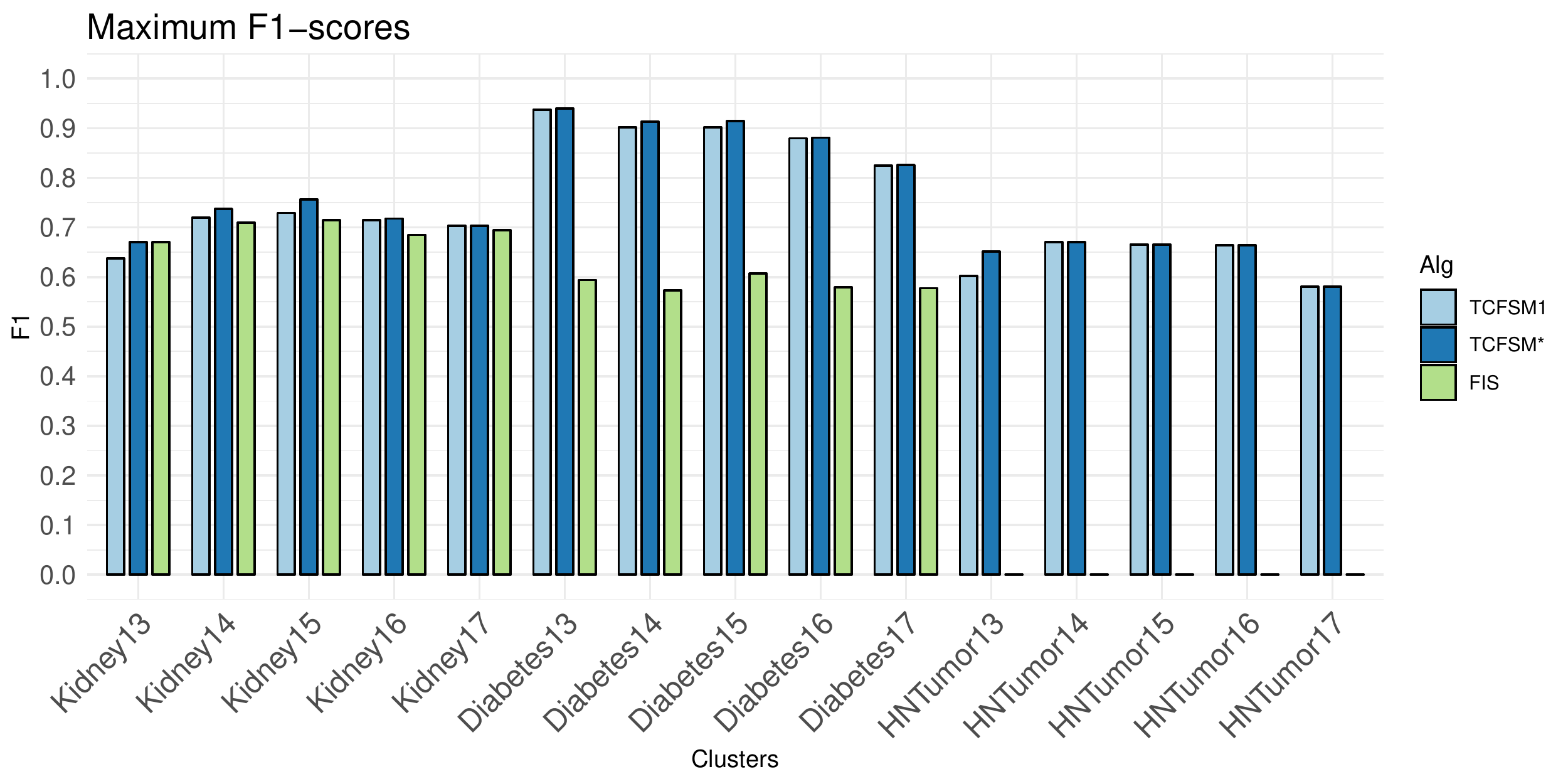}
	\caption{F1-measures achieved by our approaches TC-FSM1 and TC-FCM$^*$ for the three groups, compared to the ones achieved by the previous approach FIS~\cite{aime17amin}.}
	\label{fig:Book1}
\end{figure}

\autoref{fig:Book1} shows the maximum F1-scores of \TCb, \TCs, and \FIS on the 15 clusters over 5 years and three groups. For the diabetes group, we achieved a considerable improvement of 0.2-0.3 in the F1-scores, compared to the \FIS approach~\cite{aime17amin}. 
Overall, our approach achieved a better result. One reason for this improvement is with the use of sequential patterns (instead of frequent item sets), our approach is able to decrease the number of false positives and find the clusters with a higher precision.

\subsubsection{EO3) Frequent Sequence Patterns to Simple Process Maps.}
\begin{figure}[tbp]
	\centering
	\begin{minipage}{.5\textwidth}
		\includegraphics[width=1.00\textwidth]{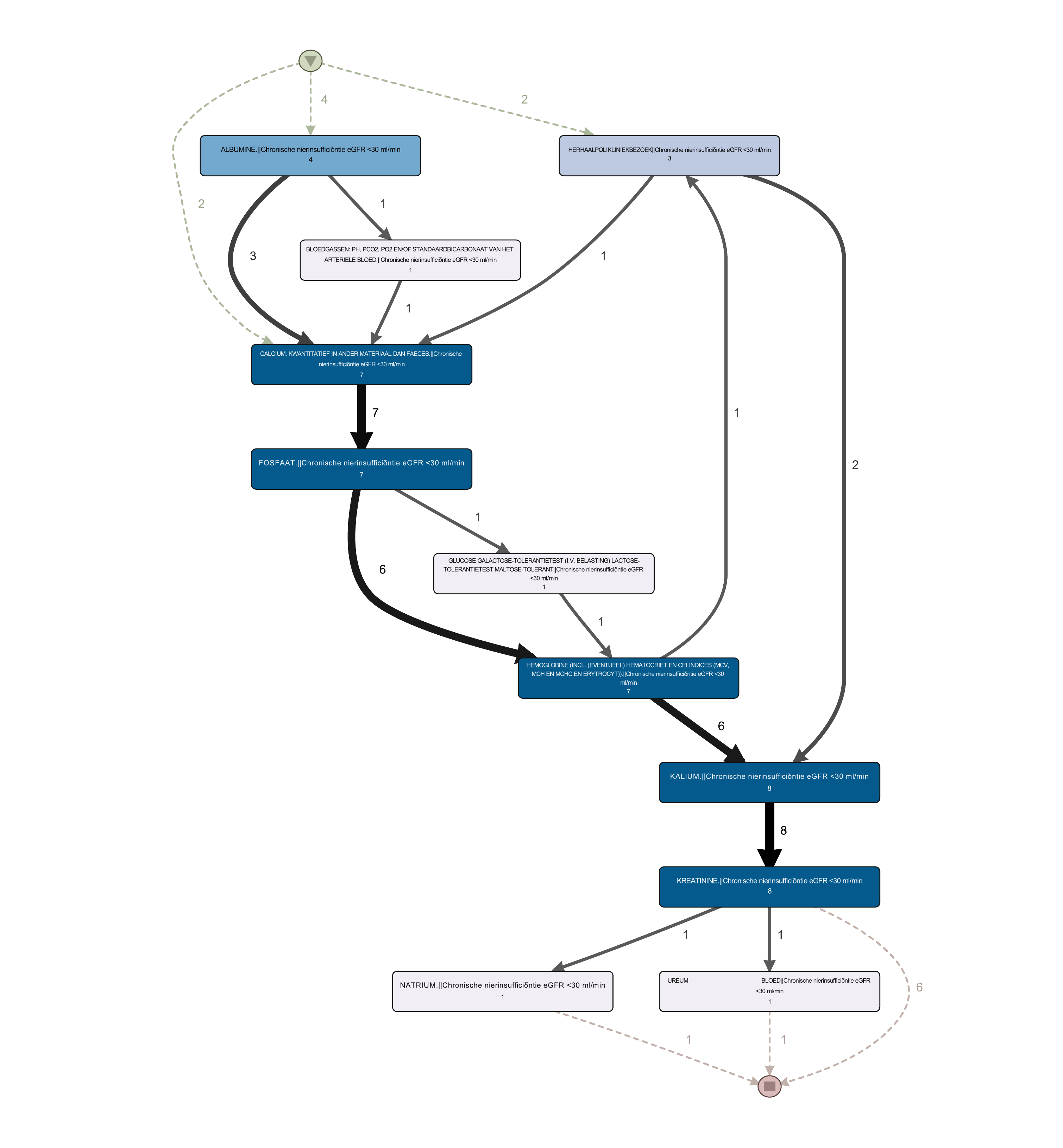}
	\captionof{figure}{The full process map based on the closed FSPs of the kidney group; ~600 activities are reduced to 11 labels.}
	\label{fig:patternlogNierschade2017k10s08}
%\end{figure}
\end{minipage}%
\begin{minipage}{.5\textwidth}
%\begin{figure}[htbp]
	\centering
		\includegraphics[width=1.00\textwidth]{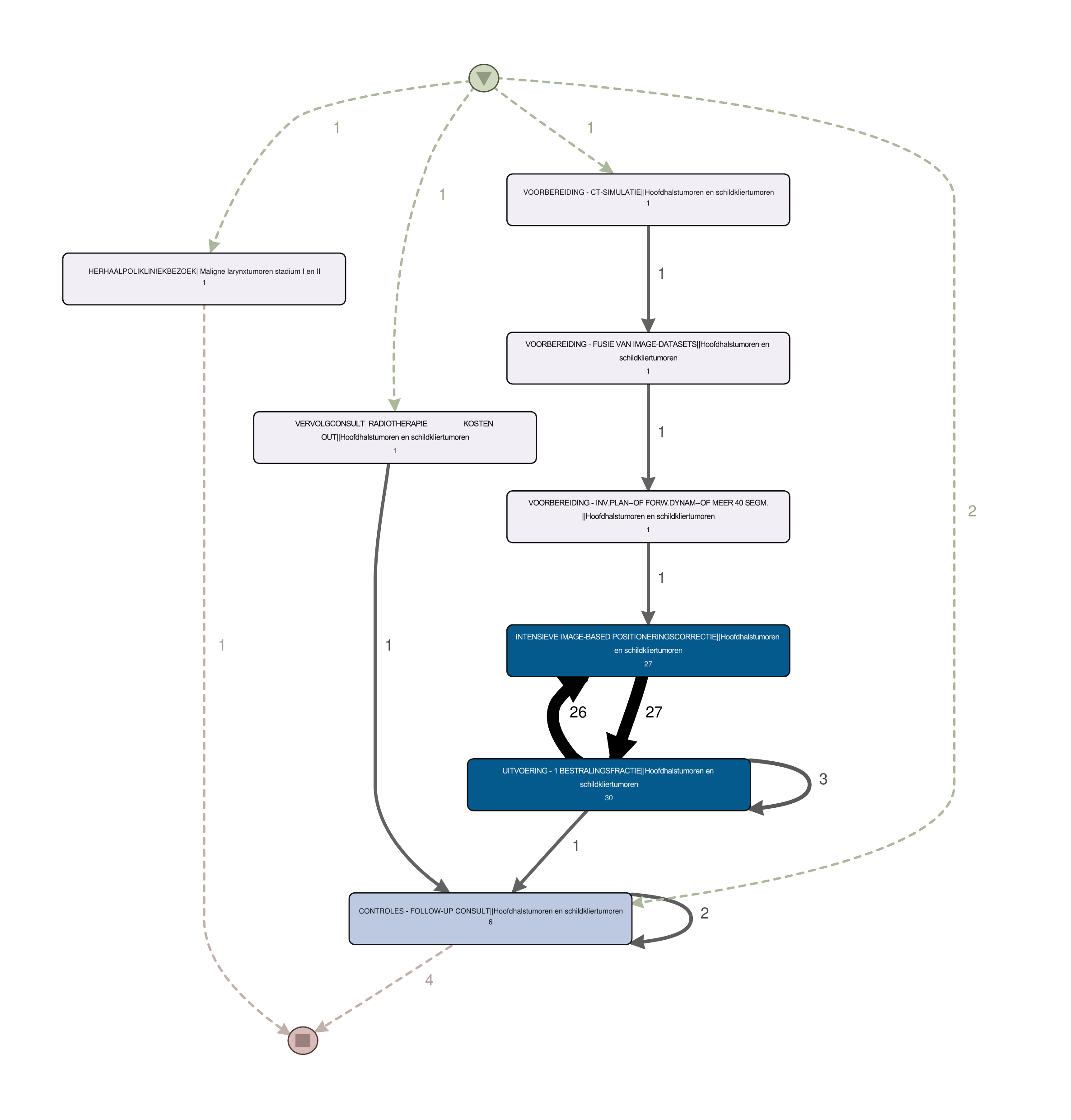}
	\captionof{figure}{The full process map based on the closed FSPs of the HNTumor group.}
	\label{fig:patternlogHNTumor2016k10s03}
	\end{minipage}
\end{figure}

\begin{figure}[htbp]
	\centering
		\includegraphics[width=0.95\textwidth]{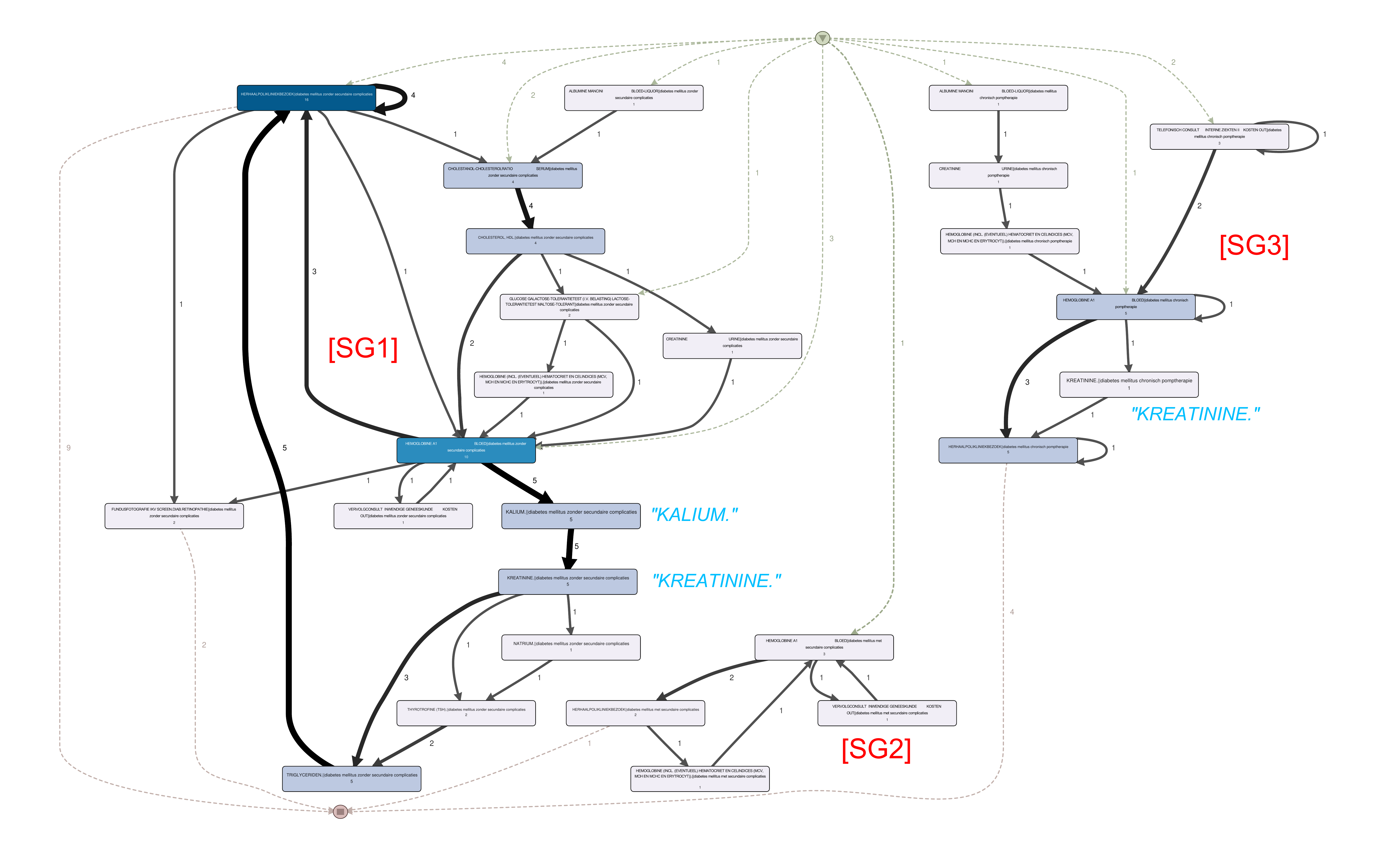}
	\captionof{figure}{The full process map based on the closed FSPs of the diabetes group; three distinct subgroups SG1, SG2, and SG3 are found.}
	\label{fig:patternlog_Diabetes2016k10s02}
\end{figure}

We used the closed sequence patterns mined on the samples as the traces that represent the frequent behavior shared by the group. Using these sequence patterns, the discovered process maps overall seem to be simple and insightful, representing only the crucial behavioral criteria of the patient groups. 
We show three process maps in~\autoref{fig:patternlogNierschade2017k10s08}, \autoref{fig:patternlogHNTumor2016k10s03} and~\autoref{fig:patternlog_Diabetes2016k10s02}, for $C_{kidney17}$, $C_{HNTumor16}$, and $C_{Diabetes16}$ to illustrate our results. All the process maps contain all activities and paths (thus no filters applied). 
As can be seen in~\autoref{fig:patternlogNierschade2017k10s08}, the number of activity labels in the process is reduced from about 4,700 to 11. The number of distinct variants is reduced from 140 to 8. 

The process maps are shown to the semi-medical expert and the data analyst. The semi-medical expert observes and confirms that the activities shown (e.g., 
\actlabel{kalium} (potassium), 
\actlabel{kreatinine} (creatinine), 
\actlabel{calcium} (calcium), 
\actlabel{fosfaat} (phosphate), 
\actlabel{albumine} (albumin), 
\actlabel{natrium} (sodium), 
\actlabel{ureum bloed} (ureum blood), etc.) are important activities (e.g., lab activities) in the clinical pathway of the kidney groups (patients with renal insufficiency). The data analyst confirms that the diagnosis code \actlabel{Chronic renal failure eGFR $<30$ ml/min} associated with these activities is a crucial criteria for defining the kidney groups.

In~\autoref{fig:patternlogHNTumor2016k10s03} and \autoref{fig:patternlog_Diabetes2016k10s02}, we also observe that multiple distinct diagnosis codes are used for the HNTumor and diabetes group, respectively. In the process map for $C_{Diabetes16}$, we found the process map being divided into three sub processes based on the diagnosis codes: 
[SG1] \actlabel{diabetes mellitus without secondary complications}, 
[SG2] \actlabel{diabetes mellitus with secondary complications}, and 
[SG3] \actlabel{diabetes mellitus chronic pump therapy}
(see \autoref{fig:patternlog_Diabetes2016k10s02}, highlighted in red).

The semi-medical expert also observes and confirms that some of these activities are important indicators for different groups. For example, \actlabel{creatinine} is important for both the kidney and diabetes groups. Nevertheless, because our approach is able to combine and handle the activities with their diagnosis codes as activity labels (in terms of the large variety of distinct labels and process variants), it enabled us to accurately distinguish the \actlabel{creatinine} for the kidney group (i.e., \actlabel{creatinine$||$Chronic renal failure eGFR $ <30 $ ml/min})
%``KREATININE.$||$Chronische nierinsufficientie eGFR $<30$ ml/min'') 
versus the same \actlabel{creatinine} but for the diabetes group (i.e., \actlabel{creatinine$||$SG1} and \actlabel{creatinine$||$SG3}, see~\autoref{fig:patternlog_Diabetes2016k10s02}, highlighted in blue).

%\subsubsection{F1 scores}
%Previous work~\cite{} used support value of 0.8

\subsection{Discussion}
The results have shown that 
%(1)
our approach using the discovered and selected frequent sequence patterns can help to cluster the patient groups with a reasonably high accuracy (e.g., a maximum of 0.75 for the Kidney group, 0.94 for Diabetes, and 0.67 for HNTumor), despite the very large data sets (on average, about 130,000 of patients and $3.9$ millions of events per year) and the relatively very small and unbalanced clusters.
Moreover, the proposed approach that automatically determines the parameters achieved F1-scores that are very close to the optimal scores. This means after setting the support $\phi_s$ and with no further input, we can find the clusters with a reasonable quality as well. 
% an average difference in F1-score of xxxx, and a maximal difference of xxx. \todo{update}
Using the process maps, we show the meaningful and insightful behavioral patterns and criteria in the clinical pathways of the patient groups. 
According to the semi-medical expert, the maps can be a useful tool in the communication with the domain experts regarding the pathways. Note that we do not have any prior knowledge of the specific activities or diagnosis codes of the patient groups. 

A remark is that these process maps of FSPs have a different semantics than the formal process models discovered using the traces. For example, an edge from A to B in the map, in essence, means that such an \term{eventually-followed} relation is frequent. 
%there is at least one closed FSP found that supports this A \term{eventually-followed} by B relation. 
To obtain formal models, we may project the patterns on the traces and use the instances of the patterns in the traces to discover models~\cite{DBLP:conf/otm/LuFASWHA17}.

%Why low 

\section{Conclusion and Future Work} 
\label{sec:conclusion}
In this paper, we investigated the trace clustering problem in healthcare and proposed an approach that can handle the characteristics of healthcare data. 
Using a small sample set of patients, the proposed approach finds frequent sequence patterns and uses these as behavioral criteria for determining a cluster.
 %to identify patient clusters with a very reasonable quality on the basis of very limited input from medical experts. 
 %We
%proposed an approach that is able to find behavioral criteria in the clinical pathways of
%patient groups based on only 
%\todo{Update previous terminologies} 
%of patient groups based on only small samples of patients provided by medical experts.
%
%Hereto, frequent item sets have been exploited, and extended to make them
%suitable for the case at hand. 
%
%The approach results in an insightful behavior definition of the patient group, while requiring only a small input from medical experts.  
%The approach is able to identify patient clusters with a very reasonable quality on the basis of very limited input from medical experts. 
%
The results of the evaluations show that the approach is able
%based on a real life data set with three groups defined by medical experts, is able 
%to cluster the groups with reasonable high F1-scores, despite the clusters to be found are very small and unbalanced. 
to identify patient clusters with a very reasonable quality on the basis of very limited input from medical experts, despite the very large data sets and the small, unbalanced clusters (ground-truth). 
The obtained behavioral criteria also led to the generation of simple process maps, where we have some first insights that these could be actually used by medical experts. 
The semi-medical expert who works closely with medical experts was able to recognize the important activities in the clinical pathways of the patient groups. 
%The model overall seems be to insightful and simple enough to be a useful tool in the communication with medical experts about the clinical pathways. %That could be useful if we want to improve clinical pathway, quality control, etc. 
Such a method may be useful to reason about clinical pathways within hospitals for the sake of process improvement or quality control.

%The frequent sequence patterns mined can be used to discover simple process maps that can be used to communicate with the experts regards the clinical pathways of these groups. 
%
%The proposed approach is very feasible since it only ask a small input from the the domain experts. 
%
%Moreover, The approach is
%
%exible in that it allows for varying strictness of the group denition by varying
%the parameters of the approach.

For future work, we plan to investigate other strategies for selecting sequence patterns as behavioral criteria to further improve the F1-scores. Also, the effect of sample size on the F1-scores is worth investigating. 
Another interesting direction is to exploit the frequent sequence patterns to discover formal process models for the clinical pathways of each cluster. 
Finally, we would like to validate the maps with medical experts and apply our approach to other patient clusters and other hospitals.
%in the process.
%For future work, we plan on extending the selection mechanism to make it
%more robust and improve the F1-scores even further. Furthermore, we want to
%apply the approach to more rare cases and involve medical experts more heavily
%in the process.

%\newsubsec{Acknowledgments.}\label{sec:Acknowledgments}
%Authors would like to thank YYYYY.
%newsubsec
%\subsubsection*{Acknowledgments.} 
{\footnotesize\newsubsec{Acknowledgments.} This research was supported by the NWO TACTICS project (628.011.004) 
and Lunet Zorg in the Netherlands. We would also like to thank the experts from the VUMC for their extremely valuable assistance and feedback in the evaluation. 
%: a collaboration between Vrije Universiteit Amsterdam, Utrecht University, 
}
%The NEXT Platform is developed and maintained by AFAS Software.

%\begin{thebibliography}{1}
%
%\bibitem{Einstein}
%A. Einstein, On the movement of small particles suspended in stationary liquids required by the molecular-kinetic theory of heat, Annalen der Physik 17, pp. 549-560, 1905.
%
%\end{thebibliography}

%\newpage

\bibliographystyle{splncs}
% argument is your BibTeX string definitions and bibliography database(s)
\bibliography{reference}

%\appendix
%\input{files/app}

\end{document}